\documentclass[journal]{IEEEtran}

\usepackage{ifpdf}
\usepackage{cite}
\usepackage{algorithmic}
\usepackage{array}
\usepackage{stfloats}
\usepackage{url}
\usepackage{amsmath,amssymb,amsfonts}
\usepackage{graphicx}
\usepackage{textcomp}
\usepackage{xcolor}
\usepackage{caption}
\usepackage{enumitem}
\usepackage{subfig}
\usepackage{breqn}
\usepackage{comment}
\usepackage{fancybox}

\begin{document}

\title{Online Long-Term Voltage Stability Margin Estimation for IBR/DER Dominated Power System with Integrated VSM-Aware TSO-DSO Framework}

\author{Ahmed Alkhonain,~\IEEEmembership{Student Member,~IEEE,}
        Kiran Kumar Challa,~\IEEEmembership{Senior Member,~IEEE,}
        Amarsagar Reddy Ramapuram Matavalam,~\IEEEmembership{Member,~IEEE,}
        Alok Kumar Bharati,~\IEEEmembership{Member,~IEEE,}
        and Venkataramana Ajjarapu,~\IEEEmembership{Life~Fellow,~IEEE}
\thanks{A. Alkhonain, K. K. Challa, and V. Ajjarapu are with the Department of Electrical and Computer Engineering, Iowa State University, Ames, IA, USA. e-mail: (ahmedkh@iastate.edu, kiranc@iastate.edu \& vajjarap@iastate.edu).}
\thanks{A. R. Ramapuram Matavalam is with the School of Electrical, Computer \& Energy Engineering, Arizona State University, Tempe, AZ, USA. E-mail: (amar.sagar@asu.edu).}
\thanks{A. K. Bharati is a Senior Researcher, Richland, WA, USA. e-mail: (alok.bharati@outlook.com).}}

\maketitle

\begin{abstract}
The rapid growth of inverter-based resources (IBRs) and distributed energy resources (DERs) has fundamentally altered the long-term voltage stability characteristics of modern power systems. This article leverages the advantages of machine learning (ML) for the online estimation of long-term voltage stability margin (VSM) and enhance the VSM though the coordinated transmission system operator–distribution system operator (TSO–DSO) optimization. An explicit analytical VSM expression is derived from offline T\&D co-simulation data using a physics-informed ML trained model under probabilistic loading and generation-mix scenarios while accounting for unbalanced distribution modeling. The resulting closed-form VSM representation is linearized and embedded into the TSO optimization problem, enabling real-time enforcement of minimum VSM constraints. We propose to enhance the operational efficiency by including the VSM sensitivities into both transmission and distribution optimization which allows the prioritization of most influential reactive power resources. Simulation studies conducted on the IEEE 30-bus transmission network integrated with multiple IEEE 37-node distribution feeders validate that the proposed framework successfully attains the desired VSM enhancement while maintaining high estimation accuracy. 
\end{abstract}

\doublebox{\parbox{0.8\linewidth}{
\centering
This paper has been submitted to IEEE Transactions on Power Systems.
}}
\\
\\


\section{Introduction}

\IEEEPARstart{T}{he} global energy landscape is undergoing a profound transformation driven by the rapid deployment of renewable generation. In the United States, electricity generation from renewable sources is projected to increase from 19\% to 38\% by 2050 \cite{ref1}, with similar transitions observed in Europe and Australia \cite{ref3, ref4}. As a result, modern power systems are evolving toward high penetrations of IBRs and DERs, fundamentally altering system dynamics, reactive power characteristics, and voltage stability behavior. Although this transition introduces operational challenges due to limited inherent reactive power support, it also presents new opportunities. Regulatory frameworks such as NERC VAR-001-6 standard, FERC Order 2222, and IEEE 1547-2018 \cite{ref5, ref6, ref7} explicitly enable and incentivize IBRs and DERs to provide reactive power support. If strategically coordinated, IBRs and DERs can become active participants in voltage stability enhancement rather than passive grid following devices \cite{ref9, ref10}. However, leveraging these capabilities requires accurate and computationally efficient methods to quantify VSM under realistic transmission–distribution interactions.

Traditional voltage stability assessment techniques, such as continuation power flow (CPF) \cite{ref11}, evaluate the maximum loadability limit primarily from a transmission-level perspective. In DER-dominant grids, this approach becomes insufficient, as voltage instability may originate from either transmission or distribution networks \cite{ref12, ref13}. Moreover, unbalanced loading and phase-specific DER injections in distribution systems significantly affect long-term voltage stability. Therefore, accurate VSM estimation in modern grids necessitates integrated transmission and distribution (T\&D) co-simulation frameworks capable of capturing unbalanced distribution characteristics. While studies such as \cite{ref9} demonstrate improved VSM accuracy using detailed T\&D modeling, these approaches rely on repetitive CPF-based simulations and lack an analytical representation of VSM, limiting their suitability for real-time TSO–DSO coordination and optimization.

To enable faster VSM estimation, researchers have explored relationships between VSM and measurable system variables. Early work \cite{ref20, ref21} established linear relationships between VSM and reactive power reserves (RPRs) of synchronous generators. Although effective in conventional generator-dominated systems, such relationships become increasingly nonlinear and complex in grids with high IBR/DER penetration, where reactive power support is geographically dispersed and electronically controlled. 

More recently, ML approaches have been proposed for real-time VSM estimation using PMU data \cite{ref22, ref23 ,ref24}. These methods improve computational efficiency and estimation accuracy. Similarly, the work in \cite{ref25} leverages stochastic modeling to construct scaled Jacobian matrices in real time. Despite their advantages, existing ML-based VSM estimation methods generally treat DERs as passive elements, overlook unbalanced distribution modeling, and do not embed VSM explicitly into coordinated TSO-DSO optimization frameworks. Consequently, while they provide fast stability assessment, they do not directly enable control-oriented VSM enhancement.

Parallel research efforts have focused on incorporating voltage stability constraints into optimal power flow (OPF) formulations. Several works \cite{r16, r17} integrate voltage stability indices into OPF. However, these indices are indirect measures and do not explicitly quantify VSM in megawatts. The work in \cite{r22} introduces an explicit VSM formulation based on reactive power reserves, but its applicability is primarily limited to synchronous generator-dominated systems. Furthermore, these approaches are restricted to transmission-level OPF and do not consider DER participation or TSO–DSO coordination.

Recent studies have emphasized the importance of TSO–DSO coordination. For example, \cite{ref9} analyzes DER impacts on VSM through T\&D co-simulation but does not incorporate centralized DER dispatch. Coordinated T\&D OPF models have been proposed in \cite{r10, r14, r15}. However, they assume balanced distribution systems, limiting realism under practical unbalanced operating conditions. The methodology in \cite{r11} utilizes distribution flexibility to support transmission voltages but lacks explicit VSM representation within the optimization. Overall, existing TSO–DSO coordination frameworks do not incorporate an explicit, computationally efficient VSM expression suitable for real-time co-optimization.

The literature therefore reveals several critical research gaps. First, there is a lack of an explicit and computationally efficient VSM expression that can be directly embedded into real-time optimization frameworks for IBR/DER-dominant systems. Most existing approaches either rely on computationally intensive CPF-based analysis or utilize indirect stability indices that do not provide a clear analytical relationship between system states and VSM. Second, limited attention has been given to modeling unbalanced distribution systems in VSM estimation, despite their significant impact on long-term voltage stability in modern grids. Third, existing TSO–DSO coordination frameworks do not explicitly incorporate VSM into the optimization problem, preventing direct control-oriented enhancement of voltage stability. Additionally, the effects of load variability and evolving generation mixes on long-term VSM have not been comprehensively investigated, particularly under high IBR/DER penetration scenarios. Finally, current methodologies do not fully exploit VSM sensitivities to identify and prioritize the most effective reactive power resources, limiting the efficiency of coordinated reactive power dispatch for stability enhancement.

To address these gaps, this paper proposes:

\begin{enumerate}
    \item An explicit and computationally efficient expression for VSM estimation is derived using a transfer learning approach within an integrated TSO-DSO co-simulation framework. The proposed model accurately captures transmission--distribution interactions and explicitly accounts for the unbalanced characteristics of distribution networks.

    \item A coordinated TSO-DSO co-optimization framework is developed that embeds the explicit ML-derived VSM expression directly into the optimization problem, enabling identification and prioritized utilization of the most VSM-sensitive synchronous generators, IBRs, and DERs.
    
    \item A distribution-level linear optimization formulation based on VSM sensitivities is proposed to dispatch DER reactive power efficiently, ensuring that the most influential DERs respond first to satisfy the aggregated reactive power request for VSM enhancement while minimizing control effort.
\end{enumerate}

The rest of the paper is organized as follows: Section II describes VSM monitoring. Section III describes the problem formulation and the TSO-DSO co-optimization framework. Section IV describes the case study results of off-line training to estimate VSM and show the impact of the proposed TSO-DSO framework. Section V concludes the paper.

\section{VSM Monitoring}

\begin{figure}[htbp]
\subfloat[Synchronous generator]{\includegraphics[width=0.25\textwidth]{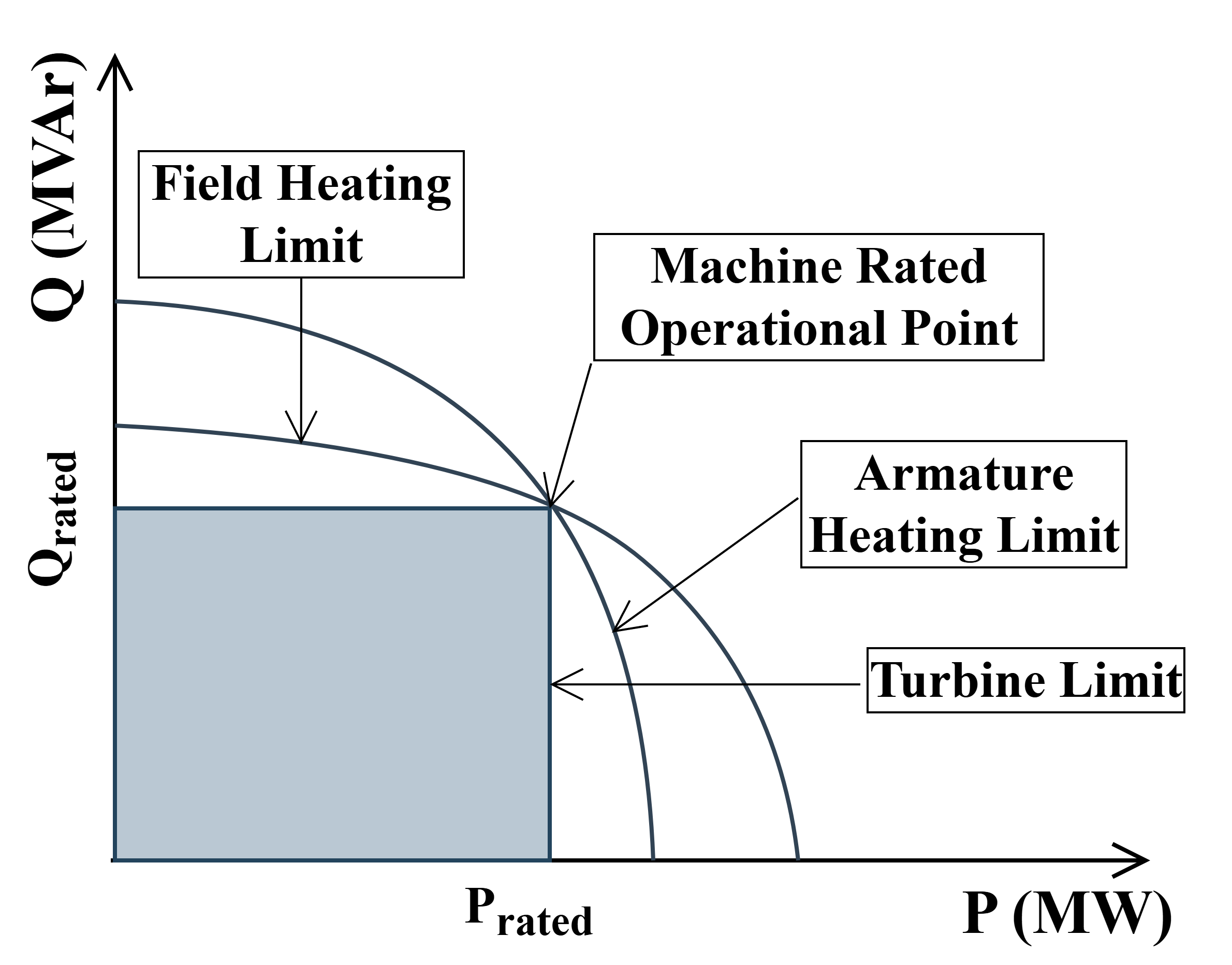}
\label{6a}}
\subfloat[DER \cite{ref30}]{\includegraphics[width=0.25\textwidth]{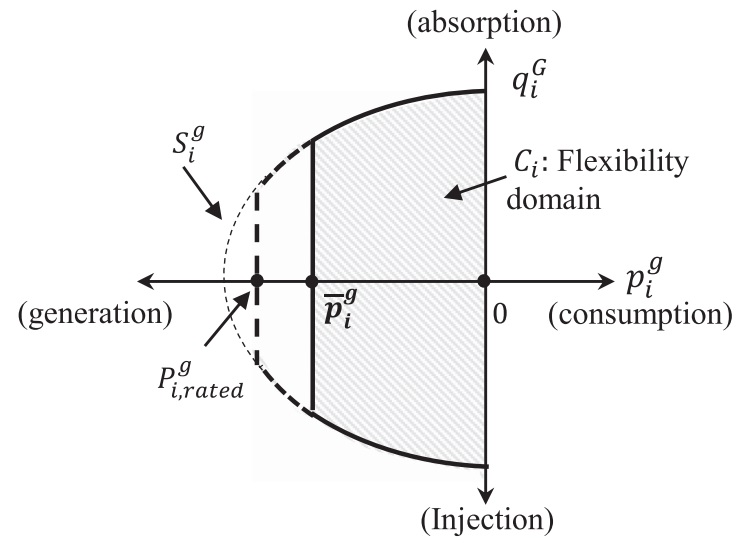}
\label{6b}}
\caption{Capability curve for synchronous generator and DERs}
\label{fig2}
\end{figure}

As discussed previously, VSM does not possess a readily available explicit analytical expression that can be directly measured online. Conventionally, VSM monitoring and control have relied on the strategic management of RPR across the power system. RRR is defined as the remaining reactive capability before reaching physical or operational limits at a given active power output \cite{ref29}. This can be directly inferred for synchronous generators from the feasibility region defined by the capability curve shown in Fig.~\ref{fig2}-(a). The available upward or downward reactive reserve corresponds to the distance from the current operating point to the nearest reactive limit. Under these assumptions, RPR can be treated as a stable and measurable indicator for VSM monitoring. The reactive power capability of IBRs, on the other hand, is constrained by their inverter continuous rating (ICR), as illustrated by IEEE IEEE 2800 Standard \cite{ref18}. At rated active power output, the minimum and maximum guaranteed reactive capability is modeled as $0.3287\cdot ICR$, representing the required reactive power absorption or injection.

In contrast, estimating RPR for DERs is significantly more challenging. DERs operate within a dynamic P--Q capability region constrained by their apparent power rating, as shown in Fig.~\ref{fig2}-(b). It is evident from Fig.~\ref{fig2}-(b) that any increase in real power generation directly reduces the available reactive capability. Consequently, reactive reserve is no longer constant but varies continuously with real-time operating conditions. Moreover, DERs are often decentralized and may operate in autonomous modes, which complicates coordinated RPR estimation. As DER penetration increases, these characteristics introduce uncertainty into reserve-based VSM monitoring. Reactive reserve does not fully capture the nonlinear interactions between transmission and distribution networks. These limitations reveal that RPR-based monitoring alone is insufficient in IBR/DER-dominant systems. Accurate representation of distributed resources in critical for the VSM assessment \cite{ref9}. Aggregating DER generation with load at the transmission level neglects feeder-level constraints and unbalanced conditions, potentially leading to an inaccurate stability assessment. Therefore, accurate VSM monitoring in modern grids requires detailed T\&D co-simulation.

Various RRR based VSM estimation and improvement methodologies are presented in the literature \cite{ref26,ref27,ref28}. The work in \cite{ref26} highlights that maintaining adequate RPR, particularly near load centers, is critical for preserving long-term voltage stability. It proposes optimal placement and management of reactive support to strengthen weak grid areas. Similarly, \cite{ref27} introduces a real-time monitoring framework that evaluates proximity to voltage instability by continuously tracking available reactive reserves. Extending this concept, \cite{ref28} develops a control strategy based on the sensitivity of VSM with respect to RPR at different buses, enabling dynamic adjustment of synchronous generator reactive outputs to maximize VSM. Collectively, these studies demonstrate that monitoring and dispatching reactive reserves can effectively enhance voltage stability in conventional systems.

An alternative and more robust approach is to monitor VSM directly through system state variables rather than indirectly through RPR. High-resolution measurements from PMUs, combined with T\&D co-simulation, allow VSM to be inferred from voltage profiles, power flows, and net loading conditions. Importantly, when T\&D co-simulation is employed, the impact of DER injections is inherently reflected in the net load seen at the transmission interface, providing a physically consistent representation of system behavior. Furthermore, VSM in modern power systems is highly sensitive not only to loading levels but also to generation mix variations. Conventional VSM assessment methods often focus primarily on peak-load conditions. However, with high IBR/DER penetration, voltage instability may arise under diverse operating scenarios. Even under similar loading conditions, different generation mixes result in significantly different voltage collapse trajectories and stability margins. This highlights that VSM cannot be characterized by a single operating condition and must instead account for variability in both load and generation composition.

\section{Problem formulation}
Formulating the VSM as an explicit mathematical function requires establishing a direct relationship between VSM and measurable system states such as generator voltage magnitudes, generator active power outputs, and load demands. Conventional VSM estimation techniques rely on CPF or iterative procedures, which do not yield an explicit analytical mapping between VSM and system variables. As system size increases, the dimensionality of state variables grows significantly, making analytical derivation intractable due to nonlinear coupling between generation dispatch, voltage control, and load variability. 

Consequently, there exists no tractable closed-form formulation that directly expresses VSM as a function of system states. This challenge motivates the use of data-driven approaches, particularly ML, to approximate the nonlinear mapping between system states ($x$) and VSM ($y$). By training ML models offline using comprehensive T\&D co-simulation datasets, an explicit analytical representation can be extracted and embedded into real-time optimization frameworks.

\begin{figure}[htbp]
\centerline{\includegraphics[width=0.5\textwidth]{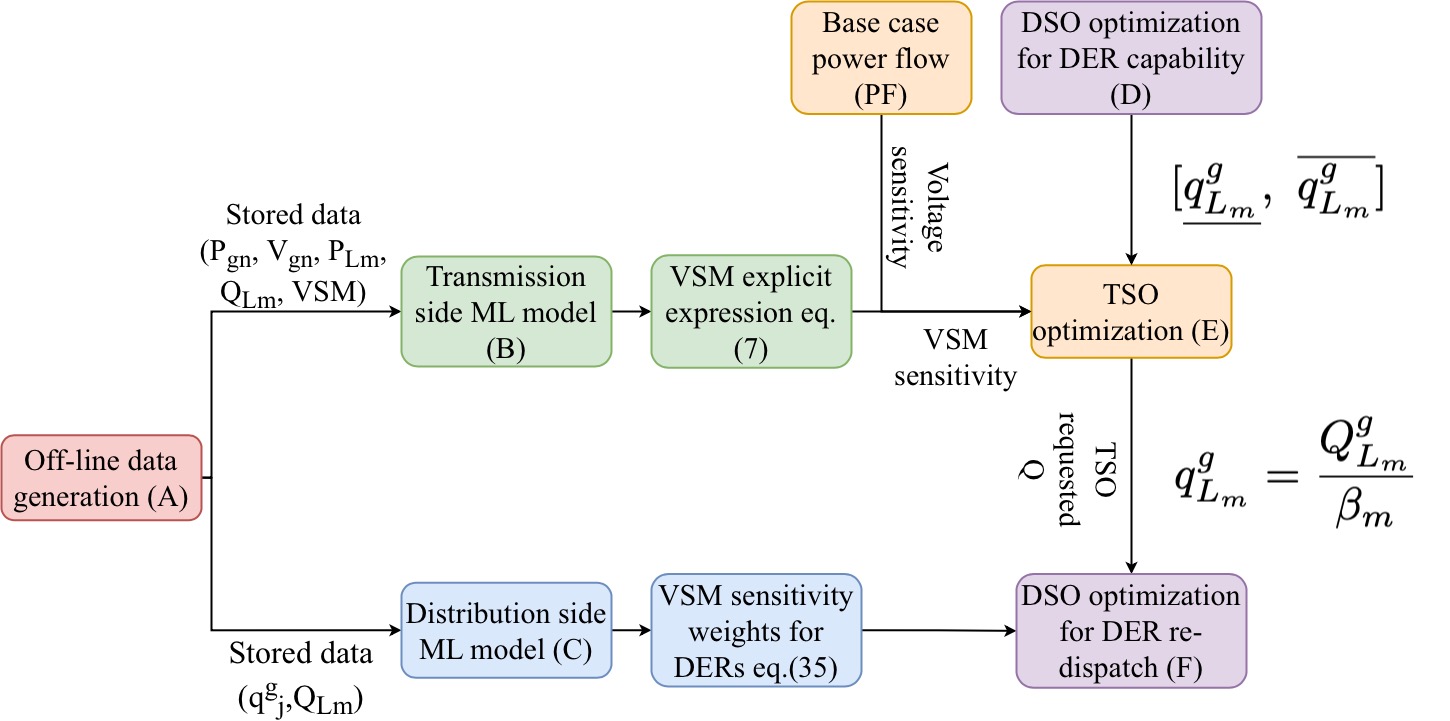}}
\caption{Proposed framework}
\label{fig6}
\end{figure}

The overall framework of the proposed methodology is illustrated in Fig.~\ref{fig6}. The approach consists of offline data generation, ML model training for both transmission and distribution systems, and integration of the explicit VSM expression into a coordinated TSO–DSO optimization structure.

The framework combines probabilistic sampling and T\&D co-simulation to generate realistic operating scenarios under high IBR/DER penetration. The trained ML models provide (i) an explicit analytical expression for VSM and (ii) sensitivity-based weights for DER reactive power dispatch. These components are then embedded into coordinated TSO–DSO optimization problems.

\begin{figure}[htbp]
\centerline{\includegraphics[width=0.5\textwidth]{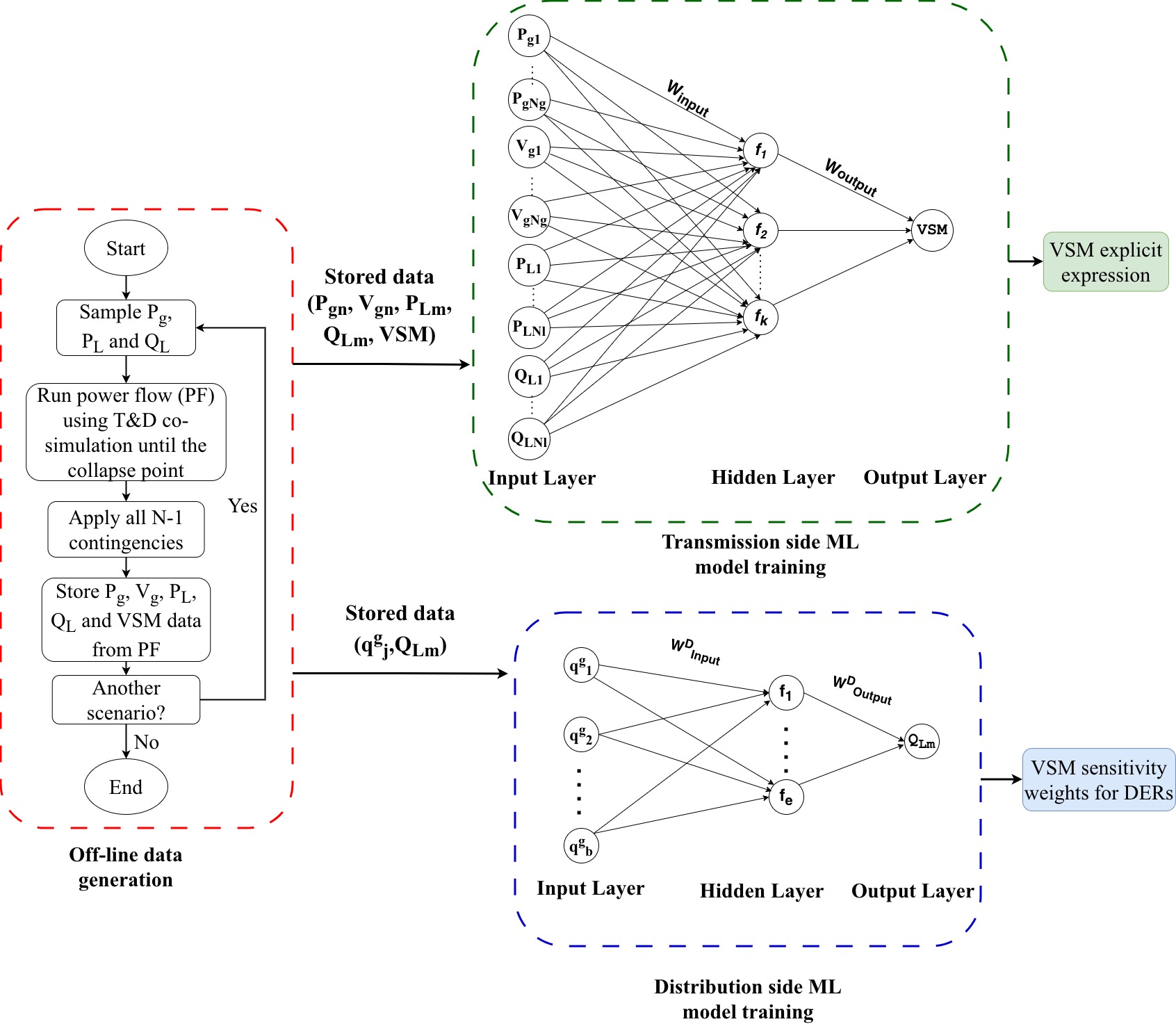}}
\caption{Proposed off-line training framework}
\label{fig5}
\end{figure}

\subsection{Scenarios definition and algorithm for VSM analysis using T\&D Co-simulation}
The proposed algorithm for VSM analysis leverages a T\&D co-simulation framework, like the work in \cite{ref32}, to accurately capture the interaction between bulk transmission and distribution networks, especially under high DER penetration. The analysis begins with the probabilistic sampling of system states. These sampled values define a unique scenario for the initial operating point for the T\&D system. Starting from this base case, the three-phase distribution power flow will run and update P\textsubscript{L} and Q\textsubscript{L} at the transmission level, then the transmission power flow will run. After that, a load increment process is applied, wherein the total system load is gradually increased using a controlled scaling factor. After each incremental step, the T\&D co-simulation is executed to solve the power flow and assess system stability. Also, all N-1 contingencies are applied for each unique sample, forming a multi-contingency for this specific scenario. Changing the sample for P\textsubscript{g}, P\textsubscript{L}, and Q\textsubscript{L} will define a new unique scenario, which will allow us to create multiple system behaviors to assess VSM.

For a one scenario, if the co-simulation successfully converges, the load is further increased, and the process is repeated. This iterative load-augmentation continues until the power flow solution fails to converge, signifying proximity to the voltage collapse point, beyond which the system cannot maintain steady-state voltage stability. The maximum loadability factor achieved before divergence defines the VSM for a particular scenario.

\subsection{Structure of transmission side ML model}
This work adopts a physics-informed sampling approach that generates data based on variability in system inputs: active power generation (P\textsubscript{g}), generator bus voltages (V\textsubscript{g}), and active/reactive load powers (P\textsubscript{L} and Q\textsubscript{L}). These inputs represent real-world grid operations, aligning with quantities managed by system operators and serving as direct inputs to the power flow equations that determine system states. Sampling these quantities ensures that the resulting voltage profiles and VSM values are physically valid and reflect realistic grid conditions.

Estimating VSM from system states constitutes a regression problem with a scalar output, specifically, a single numerical value. A commonly employed model for such regression tasks is the Multi-Layer Perceptron (MLP) neural network. The architecture of the MLP, as illustrated in Fig.~\ref{fig5}, features input variables as active power generation (P\textsubscript{g}), generator bus voltages (V\textsubscript{g}), and active/reactive load powers (P\textsubscript{L} and Q\textsubscript{L}) on the left and a single output (VSM) on the right. The intermediate nodes, called hidden neurons, incorporate nonlinear activation functions $f(.)$ that apply transformations to the input data as it propagates through the network. The activation function operates independently on each hidden neuron. The connection weights between the input and hidden layer are represented by the matrix $(W_{input})$, and those between the hidden and output layers by the vector $(W_{output})$. Although not shown for simplicity, bias terms are also present in hidden and output layers. These weights, biases, and the activation function govern the functional mapping from inputs to an output. By adjusting the parameters $W_{input}, W_{output}, b_{input}$, and $b_{output}$, the transformation from input to output can be modified.

The training involves learning the neural network parameters (weights and biases) so that the model can rapidly estimate the VSM for various unseen operating conditions through an approximate analytical mapping. This is achieved using a dataset composed of N input-output pairs $(x_1,y_1),(x_2,y_2)\cdots,(x_N,y_N)$, where $x$ is the system states and $y$ is the VSM, and formulating an optimization problem to minimize the error between the targeted VSM ($y$) and the predicted VSM ($\hat{y}$). Once the training is complete, the weight matrices and bias vectors are fixed, enabling fast and accurate VSM predictions for new system states. To train the MLP model effectively for VSM estimation Resilient Backpropagation (RProp) optimization algorithms is considered.

In this subsection, we derive the explicit mathematical expression explaining the relationship between the system states as input variables and VSM as the output variable. This explicit expression can be used to estimate the VSM at real-time operation. Let the system states be the input vector $x \in \mathbb{R}^{d\times1}$ where d is the number of inputs ($d=2n+2m$):
\begin{multline}
x^{d\times 1}= [P\textsubscript{g1}~\cdots~P\textsubscript{gn},~V\textsubscript{g1}~\cdots~V\textsubscript{gn},\\
~P\textsubscript{L1}~\cdots~P\textsubscript{Lm},~Q\textsubscript{L1}~\cdots~Q\textsubscript{Lm}]^T
\label{1}
\end{multline}
Then, the input-to-hidden weight matrix $W_{input} \in \mathbb{R}^{k\times d}$, where k is the number of neurons, is:
\begin{equation}
W^{k\times d}_{input}= \begin{bmatrix}
w_{11} &  \cdots & w_{1d}\\
\vdots &  \ddots & \vdots\\
w_{k1} &  \cdots & w_{kd}
\end{bmatrix}\label{2}
\end{equation}
Which makes the hidden layer bias vector $b_{input} \in \mathbb{R}^{k\times 1}$ is:
\begin{equation}
b^{k\times 1}_{input}= \begin{bmatrix}
b_1 & b_2 &  \cdots & b_k
\end{bmatrix}^T\label{3}
\end{equation}
The output layer weight vector $W_{output} \in \mathbb{R}^{1\times k}$ is:
\begin{equation}
W^{1\times k}_{output}= \begin{bmatrix}
C_1 & C_2 &  \cdots & C_k
\end{bmatrix}\label{4}
\end{equation}
Using the above expressions (\ref{2})-(\ref{4}), the mathematical expression for this ML model looks like the following:

\begin{equation}
\mathrm{VSM}= W^{1\times k}_{output} \cdot f(W^{k\times d}_{input} \cdot x^{d\times 1} + b^{k\times 1}_{input}) +b_{output}
\label{5}
\end{equation}
Let $h_\sigma=W_\sigma~\cdot~x+b_\sigma$, where $W_\sigma$ is the $\sigma^{th}$ row in $W^{k\times d}_{input}$, $b_\sigma$ is the $\sigma^{th}$ element in $b^{k\times 1}_{input}$, and $C_\sigma$ is the $\sigma^{th}$ element in $W^{1\times k}_{output}$.
\begin{equation}
\mathrm{VSM} = b_{output} \;+\; \sum_{\sigma=1}^{k} C_\sigma  f(h_\sigma)
\label{10}
\end{equation}

This paper considered Hyperbolic Tangent as activation functions. This activation functions is selected to describe VSM relationship with the input variable as  nonlinear activation function. The VSM expression in (\ref{10}) can be expressed as: activation function $f(h_\sigma)$ should be one of the following:

\begin{equation}
\mathrm{VSM} = b_{output} \;+\; \sum_{\sigma=1}^{k} C_\sigma \cdot tanh(h_\sigma)
\label{11}
\end{equation}

This expression in (\ref{11}) would be used to build a TSO-DSO co-optimization, like our previous work in \cite{ref33}, with the inclusion of the explicit VSM expression. This framework can enable the transmission system to enhance the long-term voltage stability using power support from IBRs/DERs in real-time.

\subsection{Structure of Distribution Side ML Model}

Similar to the transmission-side ML model developed in subsection (B), a separate neural network is constructed for the distribution system. However, instead of directly estimating the VSM, the objective of the distribution-side model is to quantify how the reactive power injections from DERs affect the net reactive power exchanged at the transmission–distribution interface. 

Specifically, the output of this model is the aggregated net reactive power at the distribution substation, denoted as $Q_{L_m}$. This quantity directly enters the transmission-side ML model as part of the system state vector and therefore indirectly influences the VSM estimation. Consequently, training the distribution-side model establishes an analytical relationship between individual DER reactive power injections and VSM through their impact on $Q_{L_m}$.

Let the input vector consist of the reactive power outputs of $b$ DERs in the distribution system:

\begin{equation}
x^D \in \mathbb{R}^{b \times 1} =
[q^{g}_{1}~\cdots~q^{g}_b]^T
\label{c1}
\end{equation}

where $q^{g}_{j}$ represents the reactive power injection of the $j^{th}$ DER. The input-to-hidden weight matrix is defined as

\begin{equation}
W^{D}_{input} \in \mathbb{R}^{e \times b} =
\begin{bmatrix}
w^D_{11} & \cdots & w^D_{1b}\\
\vdots & \ddots & \vdots\\
w^D_{e1} & \cdots & w^D_{eb}
\end{bmatrix}
\label{c2}
\end{equation}

where $e$ is the number of hidden neurons. The hidden-layer bias vector is

\begin{equation}
b^{D}_{input} \in \mathbb{R}^{e \times 1} =
\begin{bmatrix}
b^D_1 & b^D_2 & \cdots & b^D_e
\end{bmatrix}^T
\label{c3}
\end{equation}

The output-layer weight vector is

\begin{equation}
W^{D}_{output} \in \mathbb{R}^{1 \times e} =
\begin{bmatrix}
C^D_1 & C^D_2 & \cdots & C^D_e
\end{bmatrix}
\label{c4}
\end{equation}

Using (\ref{c1})–(\ref{c4}), the explicit mathematical expression of the distribution-side ML model is

\begin{equation}
\mathrm{Q_{L_m}} = b^D_{output} \;+\;  W^D_{output} \cdot tanh(W^D_{input}\cdot x^D+b^D_j)
\label{c5}
\end{equation}

This explicit expression provides a differentiable analytical mapping between DER reactive power injections and the aggregated reactive power at the substation. Since $Q_{L_m}$ is an input to the transmission-side VSM model, the sensitivity of $Q_{L_m}$ with respect to DER injections becomes essential for VSM control. The gradient of $Q_{L_m}$ with respect to the input vector $x^D$ is

\begin{equation}
\frac{\partial Q_{L_m}}{\partial x^D} = W^D_{output}\cdot \left( 1-tanh^2(W^D_{input}\cdot x^D+b^D_j)\right)
\label{c6}
\end{equation}

which provides the sensitivity weights used in the distribution-level optimization. These sensitivities quantify the effectiveness of each DER in influencing $Q_{L_m}$, and consequently, its indirect impact on VSM enhancement. Therefore, the distribution-side ML model plays a critical role in linking individual DER reactive power control actions to transmission-level voltage stability improvement within the integrated TSO–DSO co-optimization framework.

\subsection{DSO optimization formulation for DERs' reactive power capability}
 
The goal of the distribution-level optimal power flow (DSO-OPF) is to characterize the feasible range of reactive power support that DERs can provide under a load and generation profile at $T=t_0$. To derive the aggregated reactive power capability $q^g_{L_m}$, two identical OPF problems are solved: one that minimizes and another that maximizes the DERs' reactive power generation at the distribution substation, which represents the interface with the transmission system. These DSO-OPF formulations are presented and analyzed in detail in \cite{ref30}. The distribution system is modeled as a radial tree graph $\mathcal{D} = (\mathcal{N}, \mathcal{E})$, where $\mathcal{N} := \{0,1,\dots,b\}$ denotes the set of buses and $\mathcal{E} := \{(i,j)\}$ represents the set of directed branches with $i$ being the upstream node and $j$ the downstream node. For each bus $j \in \mathcal{N}$, the subset $\boldsymbol{\Gamma}_j \subseteq \mathcal{N}$ denotes the set of neighboring buses that are directly downstream of bus $j$, defined as $
\boldsymbol{\Gamma}_j := \{ \zeta \in \mathcal{N} \mid (j,\zeta) \in \mathcal{E} \}.$
 This subset includes all child nodes of bus $j$ in the radial feeder structure.

\begin{equation}
\min \; q^{g}_{L_m}
\label{ch2/1}
\end{equation}

\begin{equation}
\max \; q^{g}_{L_m}
\label{ch2/11}
\end{equation}

\noindent s.t.

\begin{equation}
P_{ij}
=
\sum_{\zeta \in \boldsymbol{\Gamma}_j} P_{j\zeta}
+
p^{l}_{j}
-
p^{g}_{j},
\quad
\forall (i,j) \in \mathcal{E}
\label{ch2/1a}
\end{equation}

\begin{equation}
Q_{ij}
=
\sum_{\zeta \in \boldsymbol{\Gamma}_j} Q_{j\zeta}
+
q^{l}_{j}
-
q^{g}_{j},
\quad
\forall (i,j) \in \mathcal{E}
\label{ch2/1b}
\end{equation}

\begin{equation}
v_j
=
v_i
-
2\left(r_{ij}P_{ij} + x_{ij}Q_{ij}\right),
\quad
\forall (i,j) \in \mathcal{E}
\label{ch2/1c}
\end{equation}

\begin{equation}
\underline{v_j}
\le
v_j
\le
\overline{v_j},
\quad
\forall j \in \mathcal{N}
\label{ch2/1d}
\end{equation}

\begin{equation}
V_{tm} \cdot \underline{\mathrm{Tap}}
\le
V_0
\le
V_{tm} \cdot \overline{\mathrm{Tap}}
\label{ch2/1e}
\end{equation}

\begin{equation}
\underline{q^{g}_{j}}
\le
q^{g}_{j}
\le
\overline{q^{g}_{j}},
\quad
\forall j \in \mathcal{N}
\label{ch2/1f}
\end{equation}

The optimization is subject to the constraints defined in Equations \eqref{ch2/1a}–\eqref{ch2/1f}. Specifically, Equations \eqref{ch2/1a}–\eqref{ch2/1c} represent the linearized distribution power flow (LinDistFlow) model, where $p^{l}_{j}$ and $q^{l}_{j}$ denote the real and reactive power demands at node $j$, respectively, $p^{g}_{j}$ is the real power output of the DER connected at node $j$, and $q^{g}_{j}$ is the DER reactive power injection, which serves as the control variable of the OPF. Equations \eqref{ch2/1d} and \eqref{ch2/1e} enforce the voltage magnitude limits at each distribution node, where $v_j=V^2_j$ and the on-load tap changer (OLTC) voltage regulation constraints, respectively. Finally, Equation \eqref{ch2/1f} imposes the inverter reactive power limits at each node $j$ hosting a DER. Solving this OPF yields the aggregated range of DER reactive power capability at the distribution substation, which represents the boundary bus interfacing with the transmission system.

\subsection{TSO optimization formulation}
This subsection presents the integration of the explicit VSM expression in \eqref{11}, obtained from the ML-based training process, into the TSO optimization framework. The proposed formulation enables coordinated utilization of the most effective reactive power resources, including synchronous generators and IBRs at the transmission level, as well as aggregated DERs at the distribution level. By embedding the explicit ML-derived VSM representation directly into the optimization problem, the framework establishes a mathematical link between reactive power control actions and system-wide voltage stability performance. Consequently, the optimization can explicitly enforce a minimum VSM requirement.

Consider a transmission system consisting of $N_T$ buses. Let 
$\mathcal{T} := \{1,2,\dots,N_T\}$ denote the set of transmission buses indexed by $\tau$. 
The system contains $N_g$ generator buses and $N_l$ load buses, defined as subsets 
$\mathcal{G} \subseteq \mathcal{T}$ and $\mathcal{L} \subseteq \mathcal{T}$, respectively. The objective function minimizes both the voltage set-point deviations at generator buses, denoted by $\Delta V_{g_n}$ for $n \in \mathcal{G}$, and the variations in reactive power support provided by distribution load buses, denoted by $\Delta Q^g_{L_m}$ for $m \in \mathcal{L}$. By limiting unnecessary reactive power adjustments, the optimization promotes smooth voltage regulation and mitigates excessive stress on generation assets.

\begin{equation}
\min 
\sum_{n \in \mathcal{G}} a_n^V (\Delta V_{g_n})^2
+
\sum_{m \in \mathcal{L}} a_m^Q (\Delta Q^g_{L_m})^2
\label{ch4/1}
\end{equation}

s.t.
\begin{multline}
\underline{V_\tau}
\le
V_\tau^0
+
\sum_{n \in \mathcal{G}}
\frac{\partial V_\tau}{\partial V_{g_n}} \Delta V_{g_n}
+
\sum_{m \in \mathcal{L}}
\frac{\partial V_\tau}{\partial Q^g_{L_m}} \Delta Q^g_{L_m}
\le
\overline{V_\tau},  \\
\forall \tau \in \mathcal{T}
\label{ch4/1c}
\end{multline}

\begin{multline}
\mathrm{VSM}^{\mathrm{current}}
+
\sum_{\sigma=1}^{k} C_\sigma
\bigg(
\sum_{n \in \mathcal{G}}
\frac{\partial f(h_\sigma)}{\partial V_{g_n}} \Delta V_{g_n}
\\+
\sum_{m \in \mathcal{L}}
\frac{\partial f(h_\sigma)}{\partial Q^g_{L_m}} \Delta Q^g_{L_m}
\bigg)
\ge
\mathrm{VSM}^{\mathrm{min}}
\label{ch4/1b}
\end{multline}

\begin{equation}
\underline{Q_{g_n}}
\le
Q_{g_n}^0
+
\frac{\partial Q_{g_n}}{\partial V_{g_n}} \Delta V_{g_n}
\le
\overline{Q_{g_n}},
\quad
\forall n \in \mathcal{G}
\label{ch4/1e}
\end{equation}

\begin{equation}
\underline{Q^g_{L_m}}
\le
\Delta Q^g_{L_m}
\le
\overline{Q^g_{L_m}},
\quad
\forall m \in \mathcal{L}
\label{ch4/1f}
\end{equation}

Where

\begin{equation}
a_n^V=1- \frac{\partial VSM /\partial V_{g_n}}{\sum_{n \in \mathcal{G}}(\partial VSM/\partial V_{g_n})}\forall n \in \mathcal{G}\label{ch4/3g}
\end{equation}

\begin{equation}
a_m^Q=1- \frac{\partial VSM /\partial Q_{L_m}}{\sum_{m \in \mathcal{L}}(\partial VSM/\partial Q_{L_m})}\forall m \in \mathcal{L}\label{ch4/3h}
\end{equation}

The objective terms are weighted by sensitivity-based coefficients $a_n^V$ and $a_m^Q$, derived from the VSM sensitivities in \eqref{ch4/3g} and \eqref{ch4/3h}, respectively. These parameters prioritize the most influential control variables in improving voltage stability. The formulation is subject to operational voltage constraints in \eqref{ch4/1c}, ensuring that all transmission bus voltages remain within acceptable limits, as well as a minimum VSM constraint in \eqref{ch4/1b}. The VSM constraint is linearized using sensitivity factors obtained from the explicit ML-based VSM model in \eqref{11}, providing an accurate yet computationally efficient representation of system stability. Reactive power capability limits of synchronous generators and IBRs are enforced in \eqref{ch4/1e}, while the flexibility of DERs is incorporated through \eqref{ch4/1f}.

It is important to clarify the definition of the reactive power variable at the transmission–distribution interface. The term $Q_{L_m}$ in \eqref{1} represents the net reactive load at transmission load bus $m$, defined as
\begin{equation}
Q_{L_m} = Q^{l}_{L_m} - Q^{g}_{L_m},
\end{equation}
where $Q^{l}_{L_m}$ denotes the aggregated reactive power demand of feeder $m$, and $Q^{g}_{L_m}$ represents the reactive power support provided by the corresponding distribution system. Therefore, increasing $Q^{g}_{L_m}$ effectively reduces the net reactive load observed by the transmission system. Furthermore, the reactive power contribution $Q^{g}_{L_m}$ appearing in the TSO optimization is a scaled representation of the distribution-level DER dispatch variable $q^{g}_{L_m}$ obtained from the DSO optimization stage. The relationship between the two levels is expressed as
\begin{equation}
Q^{g}_{L_m} = \beta_m \, q^{g}_{L_m},
\end{equation}
where $\beta_m$ is the scaling factor that maps the equivalent distribution feeder injection to the transmission base. This formulation ensures consistency between the TSO and DSO optimization problems and establishes a coordinated multi-level reactive power control framework.

\subsection{DSO optimization formulation for reactive power re-dispatch}

The final stage of the proposed framework determines the reactive power set-points of individual DERs to ensure that the DSO satisfies the reactive power request issued by the TSO. The objective function is to minimize the weighted reactive power generation of DERs while satisfying constraints in \eqref{ch2/3a}-\eqref{ch2/3c}.

\begin{equation}
\min 
\sum_{j \in \mathcal{N}} w_j q^{g}_{j}
\label{ch2/3}
\end{equation}

\noindent s.t.

\begin{equation}
\eqref{ch2/1a} - \eqref{ch2/1f},
\quad
\forall (i,j) \in \mathcal{E}
\label{ch2/3a}
\end{equation}

\begin{multline}
-\alpha \cdot q^{\mathrm{avg}}_j
\le
q^{\phi}_j - q^{\mathrm{avg}}_j
\le
\alpha \cdot q^{\mathrm{avg}}_j,\\
\quad
\forall j \in \mathcal{N}, \forall \phi \in \{a,b,c\}
\label{ch2/3b}
\end{multline}

\begin{equation}
q^{g}_{L_m}
=
\frac{Q^g_{L_m}}{\beta_m}
\label{ch2/3c}
\end{equation}

where
\begin{equation}
w_{j}=1- \frac{\partial Q_{L_m}/\partial q^{g}_{j}}{\sum(\partial Q_{L_m}/\partial q^{g}_{j})}  \forall j \in \mathcal{N}\label{ch2/3d}
\end{equation}

The three-phase radial power flow constraints in \eqref{ch2/1a}-\eqref{ch2/1f} is represented in \eqref{ch2/3a}. Equation \eqref{ch2/3b} limits the allowable increase in phase unbalance, where $\alpha$ represents the maximum permissible percentage deviation from the average three phase reactive power. Equation \eqref{ch2/3c} ensures that the aggregated reactive power contribution of the distribution system meets the scaled reactive power request from the TSO, where $Q^g_{L_m}$ is the scalar reactive power request determined by the transmission-level optimization and scaled down using $\beta_m$ which represent the number of parallel distribution systems connected to bus $m$. Unlike conventional DER dispatch approaches, each DER located at bus $j \in \mathcal{N}$ is assigned a weighting factor $w_j$ that reflects the sensitivity of the distribution substation reactive load $Q_{L_m}$ to reactive power injections at the associated DER bus $j$. These sensitivities are derived from \eqref{c6}. The weights prioritize the most influential DERs in enhancing $Q_{L_m}$, thereby reducing unnecessary participation of less effective units.
\section{Case study}
In this section, simulation results from PSSE for the IEEE 30 bus system with its loads replaced by multiple 37-node distribution systems simulated using OpenDSS. First, the test system is described. Then, the ML model training for learning the VSM function and the accuracy for estimating the VSM are described. Then, the statistical validation for the ML-trained model is described. Next, a case of TSO-DSO co-optimization that shows the most effective IBRs and DERs to enhance the VSM.
\subsection{Test system}
\begin{figure}[htbp]
\subfloat[IEEE 30 bus system]{\includegraphics[width=0.25\textwidth]{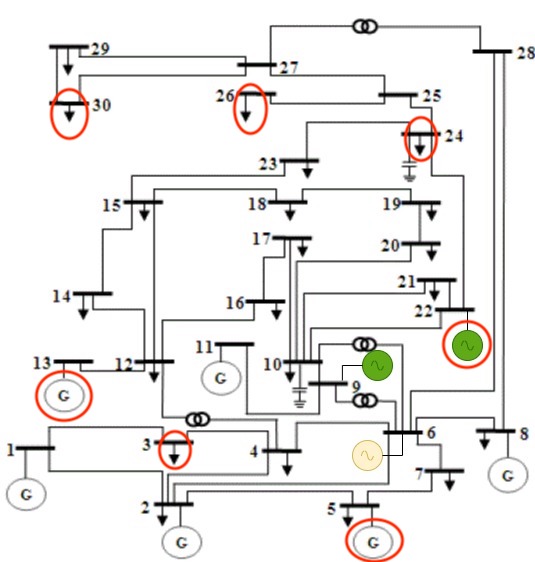}}
\label{fig7a}
\subfloat[IEEE 37 node system]{\includegraphics[width=0.23\textwidth]{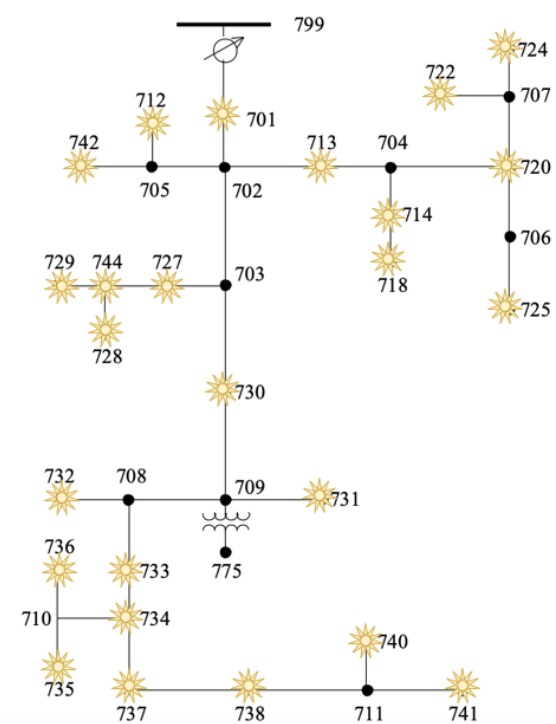}}
\label{fig7b}
\caption{Test system}
\label{fig7}
\end{figure}

The transmission system under study is the IEEE 30 bus system \cite{ref34}. This system has 6 generators and 3 IBRs connected to bus 6 (Solar PV), bus 9 and bus 22 (Wind) as shown in Fig.\ref{fig7}-(a). This system has 20 loads with a total base load of 280 MW. All the loads in the transmission system are represented by multiple parallel feeders of the IEEE 37-node distribution system \cite{ref35}. The distribution system is modified to include solar PVs as DERs. The total penetration of DERs in the distribution system as a percentage of the load is 50\%. The IBRs in the transmission system are rated to cover 10\% of the total load.

In this study, T\&D co-simulation is employed to generate PV curves under various load increase directions, capturing the system’s long-term voltage stability behavior. The VSM dataset is driven by projected wind and solar PV generation scenarios from NREL \cite{a2}, which distinguish between wind-based IBRs and PV IBR connected at the transmission level due to their different active power profiles. To ensure robustness and reflect real-world grid vulnerabilities, the simulation is systematically repeated for all possible N-1 contingency scenarios, covering the loss of individual transmission elements. Each simulation yields the system state variables along with the resulting VSM. This process generates a dataset comprising approximately 3,800 unique data points, with each data point representing system states paired with its corresponding VSM. This dataset forms the foundation for training and validating the ML model used for VSM estimation.

\subsection{Statistical Validation of VSM Expression using ML}

The VSM expression using ML-trained model, which was mentioned in section III-B, is evaluated based on its prediction accuracy, measured by the coefficient of determination ($R^2$), which quantifies the proportion of the variance in the dependent variable that is predictable from the independent variables, indicating how well the predicted values approximate the actual data, where 1 is the ideal value. Also, the mean absolute error (MAE) is used, which measures the average magnitude of the errors between predicted and actual values. MAE is in the unit of MW, so to make it more understandable, a standardized MAE is used where the error is divided by the actual VSM. The results indicate that the VSM expression can capture the nonlinear relationship between system states and VSM. The ML-trained VSM expression achieved a high $R^2$ of 0.969 and MAE of 4.36\%, demonstrating superior generalization and robustness.

Finally, a 5-fold cross-validation (CV) procedure was employed to evaluate the model’s performance and robustness. This method divides the dataset into five equal parts, sequentially using four parts for training and one for testing, ensuring that every data point is used for training and validation. The results from the 5-fold CV yielded an MAE of 4.4\%, and  $R^2$ of 0.9574. These metrics indicate that the model consistently produces accurate predictions with low average error across different data splits. The relatively high $R^2$ value further confirms that the model explains a significant portion of the variance in the data, demonstrating strong predictive capability and reliability.

\subsection{Scalability of The Proposed real-time VSM estimation}
To evaluate the scalability of the proposed ML-based VSM estimation, the model was applied to two systems of varying sizes and complexity: the IEEE 30-bus system and the IEEE 118-bus system. Similar to the case of the IEEE30-bus system, each load in the IEEE 118-bus system is replaced with a parallel IEEE 37-node test system for the T\&D co-simulation. This comparison aims to determine how well the method maintains accuracy and computational efficiency as the system size increases. This is critical for assessing the model’s suitability for real-world applications involving large-scale T\&D networks with high IBR/DER penetration.

The two models were trained and tested using the same ML architecture, Resilient Backpropagation with Tanh as activation function, with adjustments only to accommodate the dimensionality of the input vectors reflecting system states. The performance metrics in Table~\ref{ch3/tab4} across the two systems demonstrate strong consistency. Specifically, the MAE as a percentage of the actual VSM values remained nearly identical between the 30-bus and 118-bus systems, indicating that the model maintains high predictive accuracy regardless of system size. Similarly, the $R^2$ showed no significant degradation in performance, further validating the generalization capability of the ML across different network scales.

These results confirm that the proposed ML-based framework exhibits strong scalability in accuracy and efficiency. Making it a viable solution for large-scale power systems with complex T\&D interactions and high IBR/DER integration.

\begin{table}[htbp]
    \caption{Scalability Comparison}
    \begin{center}
        \begin{tabular}{|c|c|c|}
            \hline
            \textbf{System} & \textbf{MAE \%} & \textbf{R\textsuperscript{2}}                                                        \\
            \hline
            \textit{IEEE-30 bus} & 4.36 & 0.969 
            \\
            \hline
            \textit{IEEE-118 bus} & 4.4 & 0.95 \\
            \hline
        \end{tabular}
        \label{ch3/tab4}
    \end{center}
\end{table}

\subsection{TSO Objective Function Parameters Based on VSM Sensitivities}

Starting from a stressed operating condition at total load 750 MW with an initial VSM before control (optimization) of 13.08 MW, the transmission-side optimization re-dispatches reactive power resources to achieve the targeted minimum VSM of 90 MW. The proposed formulation incorporates VSM sensitivity-based weighting parameters into the objective function in order to prioritize the most effective control variables for stability enhancement.

The coefficients $a_n^V$ and $a_m^Q$ reported in Tables II and III in a ascending order, where the low coefficient means more effective reactive power source, are computed using \eqref{ch4/3g} and \eqref{ch4/3h}, respectively. These expressions are derived from the partial derivatives of the explicit VSM formulation in \eqref{11}. Specifically, the sensitivities are obtained by evaluating the derivatives of the activation function terms $f(h_\sigma)$ with respect to the corresponding control variables. The inputs to $f(h_\sigma)$ are obtained from the power flow solution at the current loading condition prior to applying any control actions. This ensures that the weighting factors accurately reflect the real-time operating point of the system. These coefficients reflect the marginal contribution of each control variable to VSM improvement. The non-uniform distribution of these parameters clearly indicates that certain buses  provides greater influence on voltage stability compared to others. 

\begin{table}[htbp]
\caption {Generators and IBRs VSM Sensitivity Coefficients}
\label{table1}
\centering
\begin{tabular}{|c|c|c|c|c|c|c|c|c|}
\hline
Bus No. & 5 & 22 & 13 & 11 & 9 & 2 &  8 & 6 \\
\hline
$a^v_{n}$ & 0.53 & 0.64 & 0.65 & 0.73 & 0.82 & 0.995 & 0.98 & 0.986 \\
\hline
\end{tabular}
\end{table}

\begin{table}[htbp]
\caption {DERs VSM Sensitivity Coefficients}
\label{table2}
\centering
\begin{tabular}{|c|c|c|c|c|c|c|c|c|}
\hline
Bus No. & 30 & 26 & 24 & 3 & 29  & 21 & 12 & 10 \\
\hline
$a^Q_{m}$ & 0.25 & 0.32 & 0.34 & 0.42 & 0.53 & 0.74 & 0.83 & 0.89 \\
\hline
\end{tabular}
\end{table}

When equal weighting parameters are assigned to all control variables, the contributions come from 6 synchronous generators/IBRs and 14 aggregated DER units at the transmission level. This correspond to the set of control variables that satisfy constraint \eqref{ch4/1b} while minimizing the objective function in \eqref{ch4/1}, thereby representing the optimal reactive power resources selected at this condition. Although this approach guarantees optimality, it increases coordination requirements and expands the number of control actions that must be implemented in real time. By contrast, incorporating VSM sensitivity-based parameters allows the optimization to naturally favor the most influential controllers to improve VSM. As a result, the number of actively dispatched units is reduced to only 7 controllers, 3 synchronous generators/IBRs and 4 aggregated DERs, while still satisfying the minimum VSM requirement. The location of these 7 controllers with the corresponding reactive power and voltage adjustments are presented in Tables \ref{gen_dispatch_ieee} and \ref{tab:der_dispatch_ieee} and circled in Fig.~\ref{fig7}-a. The location of these controllers plays a critical role, as they are electrically close to major load centers that typically correspond to weak buses in the system. Under an N-1 contingency, the system operating condition and weak bus locations may shift, which is directly reflected in the VSM sensitivity coefficients. Consequently, the set of participating controllers selected by the optimization will vary depending on the specific contingency scenario.

\begin{table}[htbp]
\caption{Generators and IBRs Voltage Set-point Dispatch}
\label{gen_dispatch_ieee}
\centering
\begin{tabular}{|c|c|c|c|}
\hline
Bus No. & 5 & 22  & 13 \\
\hline
$\Delta V_{g}$ & 0.0260 & 0.02 & 0.017\\
\hline
\end{tabular}
\end{table}

\begin{table}[htbp]
\caption{DER Reactive Power Dispatch}
\label{tab:der_dispatch_ieee}
\centering
\begin{tabular}{|c|c|c|c|c|}
\hline
Bus No. & 30 & 26 & 24 & 3 \\
\hline
$\Delta Q^g_{L}$ & 6.40 & 3.84 & 2.60 & 1.00 \\
\hline
\end{tabular}
\end{table}

Each aggregated DER unit typically represents at least 20 distributed inverters at the distribution level. Therefore, limiting the number of active aggregated DER controllers substantially reduces downstream communication, coordination, and implementation complexity. The benefit of sensitivity-based controller reduction becomes even more significant in larger systems, where the number of controllable resources can grow substantially. Engaging all available controllers may increase computational burden and communication overhead without proportional improvement in voltage stability margin. The proposed sensitivity-based parameterization provides a scalable mechanism that selectively activates the most effective controllers for real-time TSO–DSO coordinated voltage stability control.

\subsection{DERs VSM sensitivity-based re-dispatch}

The proposed reactive power dispatch algorithm with the VSM sensitivity in equation \eqref{ch2/3d} is compared with the dispatch using equal weights. The VSM results after single-iteration of TSO-DSO co-simulation in table \ref{tab:vsm_single_iter_ieee} indicate that sensitivity-based dispatch strategy yield more accurate VSM estimates compared to the equally weighted approach, with the VSM sensitivity method providing the closest approximation to the target margin (90 MW). This difference from the targeted margin to the calculated VSM in table \ref{tab:vsm_single_iter_ieee} represent the error between the linear TSO-DSO co-optimization and the non-linear VSM estimation using \eqref{11}. To adjust this error, multiple rounds of TSO-DSO co-optimization is done to reach the targeted minimum VSM. Furthermore, as shown in table \ref{tab:tso_dso_iterations_ieee}, incorporating VSM sensitivities into the TSO–DSO co-optimization framework substantially reduces both the required reactive power request (from bus 24) and the number of iterations needed for convergence. The VSM sensitivity-guided dispatch achieves the desired VSM in only two iterations, compared to seven iterations for the equally weighted case. These results confirm that embedding ML-derived voltage stability sensitivities into the optimization process enables faster convergence, reduced control effort, and more efficient utilization of DER reactive power flexibility in TSO–DSO coordinated operation.

\begin{table}[htbp]
\caption{VSM Estimation After a Single T\&D Co-Simulation Dispatch Iteration}
\label{tab:vsm_single_iter_ieee}
\centering
\begin{tabular}{|c|c|c|}
\hline
Dispatch Method & Equal Weights & VSM Sensitivity \\
\hline
VSM (MW) & 79.63 & 87.30 \\
\hline
\end{tabular}
\end{table}

\begin{table}[htbp]
\caption{TSO--DSO Optimization Performance to Reach Targeted VSM (90 MW)}
\label{tab:tso_dso_iterations_ieee}
\centering
\begin{tabular}{|l|c|c|}
\hline
 & Equal Weight & VSM Sensitivity \\
\hline
Requested MVAr & 3.10 & 2.72 \\
\hline
No. of iterations & 7 & 2 \\
\hline
\end{tabular}
\end{table}

\begin{figure}[htbp]
\centerline{\includegraphics[width=0.5\textwidth]{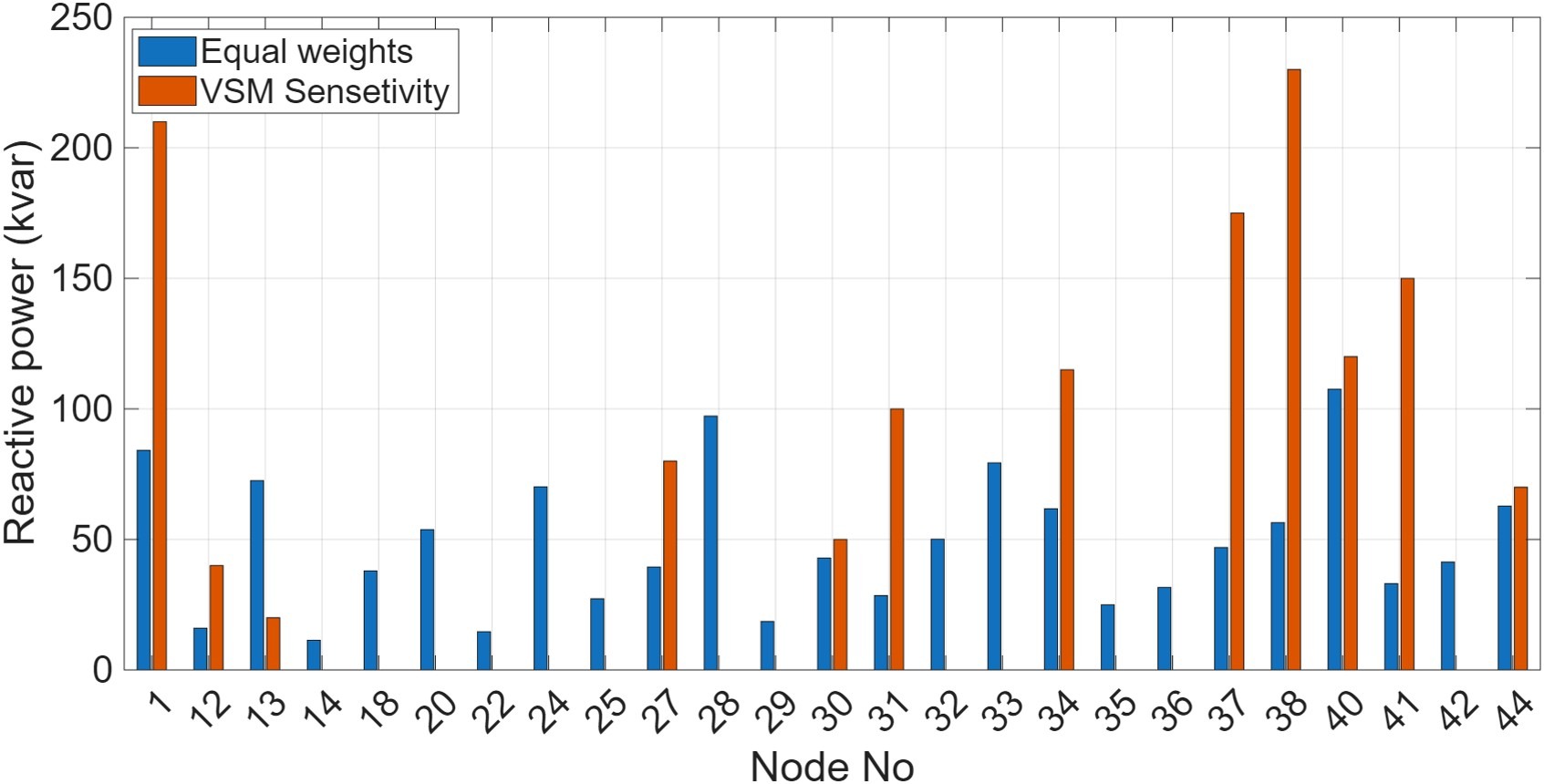}}
\caption{Distribution system dispatch
}
\label{fig12}
\end{figure}

Fig.~\ref{fig12} illustrates the DER reactive power re-dispatch across multiple nodes under the different coordination strategies. The results highlight clear differences in how reactive power support is spatially allocated when voltage stability considerations are explicitly incorporated into the optimization framework. In the equally weighted case, reactive power injections are relatively moderate and distributed without clear prioritization of critical locations, resulting in a less effective contribution to overall voltage stability improvement. While this approach provides some level of voltage support, it does not sufficiently target buses that have a strong impact on the system VSM.
In contrast, the VSM sensitivity-based dispatch exhibits a more selective and impactful allocation of reactive power, with significantly larger injections at nodes that contribute most effectively to increasing the voltage stability margin. The pronounced reactive power support at strategically important buses demonstrates that the optimization successfully identifies locations where DER flexibility yields the highest marginal VSM improvement. This targeted re-dispatch leads to a more efficient utilization of DER reactive power capability, achieving higher VSM enhancement with reduced overall control effort.

\section{Conclusion}
This paper presented an integrated ML-enabled TSO–DSO coordination framework for real-time voltage stability margin estimation and enhancement in power systems with high penetrations of DERs and IBRs. By leveraging probabilistic scenario generation and unbalanced T\&D co-simulation, an explicit analytical VSM expression was derived using a ML model trained offline on physics-consistent operating points. Unlike conventional CPF-based approaches or indirect reserve-based indicators, the proposed formulation enables direct embedding of VSM constraints into the transmission-level optimization problem, transforming VSM into a controllable operational constraint.

The results demonstrate that the ML-derived VSM expression accurately captures nonlinear transmission–distribution interactions and enables real-time evaluation. Incorporating VSM sensitivities into the TSO objective function allows selective prioritization of the most influential reactive power resources. This sensitivity-based parameterization reduces the number of participating controllers from 20 to 7. Moreover, the coordinated framework significantly reduces the number of TSO–DSO iterations required to reach the target VSM, thereby lowering communication overhead and computational burden. Since each aggregated DER represents multiple downstream inverters, limiting the number of active aggregated controllers substantially reduces implementation complexity, an advantage that becomes increasingly significant in large-scale systems with extensive DER penetration. Overall, the proposed framework establishes a scalable and practical methodology for stability-oriented reactive power coordination in modern grids.

\end{document}